\documentclass[preprint,aps]{revtex4}
\usepackage{graphicx}
\usepackage{dcolumn}
\usepackage{bm}

\begin{document}

\preprint{APS/123-QED}

\title{Room temperature spontaneous emission enhancement from quantum dots in photonic crystal slab cavities in the telecommunications C-band}

\author{R. Hostein}
\author{R. Braive}
\author{M. Larqu$\acute{\text{e}}$}
\author{K.-H. Lee}
\author{A. Talneau}
\author{L. Le Gratiet}
\author{I. Robert-Philip}
\author{I. Sagnes}
\author{A. Beveratos}
\affiliation{
Laboratoire de Photonique et Nanostructures LPN-CNRS UPR-20 \\
Route de Nozay. 91460 Marcoussis, France }

\date{\today}

\begin{abstract}
We report on the control of the spontaneous emission dynamics from
InAsP self-assembled quantum dots emitting in the
telecommunications C-band and weakly coupled to the mode of a
double heterostructure cavity etched on a suspended InP membrane
at room temperature. The quality factor of the cavity mode is
44$\times 10^3$ with an ultra-low modal volume of the order of 1.2
$(\lambda/n)^3$, inducing an enhancement of the spontaneous
emission rate of up a factor of 2.8 at 300 K.
\end{abstract}

 \maketitle

The spontaneous emission dynamics of a single emitting dipole
results from the interaction between the emitting dipole and its
electromagnetic environment. Consequently, the modification of the
electromagnetic field around the point emitter, by use of a
microcavity for instance, alters the spontaneous emission lifetime
of the emitting dipole, inducing in the weak coupling regime
either an enhancement or an inhibition of the spontaneous emission
rates \cite{Purcell}. First observed in atom physics
\cite{Goy1983, Hulet1985}, the transposition of these effects to
semiconductor physics has already been demonstrated by embedding,
for instance, self-assembled quantum dots in micropillars
\cite{Gerard1998, Bayer2001, Solomon2001}, microdisks
\cite{Gerard1999, Kiraz2001, Fang2002} and more recently photonic
crystal slab cavities \cite{Kress2005, Englund2005, Badolato2005}.
Such spontaneous emission modification can be used to increase the
modulation bandwith of nanolasers \cite{Atlug2006} or the quantum
efficiency of single photon sources for quantum communication
\cite{Barnes2002}. However, most of these experimental
demonstrations involve InAs/GaAs quantum dots emitting below 1050 nm and at low temperature, while such
prospective applications require operation in the
telecommunication wavelength range and at room temperature.
Recently, spontaneous emission enhancement by a factor of 8 has
been observed at low temperature on InAs/GaAs dots emitting in the
telecommunication O-band ($1260-1360$ nm) \cite{Balet2007}. The
remaining concern is room temperature operation. At room
temperature, spontaneous emission enhancement is
limited by the large homogeneous linewidth of emitting dipoles but
may be increased in a simple manner by reducing the modal volume
of the cavity mode, which is a particular feature of photonic
crystals. Up to now, most experiment aiming at observing a Purcell
effect at room temperature by use of photonic crystals were based
on an analysis of the amplitude of the emitted signal, in order to
infer an estimated value of the Purcell
factor \cite{Nozaki2007, Makarova2008}. Yet, direct lifetime
measurements solely allow to unambiguously determine the amplitude
of the spontaneous emission exaltation induced by the presence of
the microcavity. In this paper, we report on lifetime measurements indicating a spontaneous emission enhancement by a factor of 2.8 at room temperature on an ensemble of quantum dots emitting at 1538 nm.

Samples are grown by metalorganic vapour phase epitaxy. The InP
membrane is grown on a GaInAs sacrificial layer lying on a InP
buffer. It incorporates in its center a single layer of
self-assembled InAsP quantum dots. The quantum dots density is of
the order of 15x$10^9$ cm$^{-2}$ and their photoluminescence is
centered around 1560 nm at 300 K, with a inhomogeneous spectral
broadening of the order of 150 nm \cite{Michon2008}. After the
deposit of a SiN layer on top of the semiconductor, the cavity
pattern is defined on a layer of polymethylmethacrylate (PMMA)
using direct-write electron-beam lithography. The cavity is formed
by a double heterostructure resonator \cite{Akahane2003}. It
consists of a W1 waveguide composed of one missing row of holes in
the $\Gamma K$ direction of a hexagonal lattice structure, with a
local enhancement of the lattice period over two periods at the
center of the photonic crystal waveguide. The lattice period is
only modified along the $\Gamma K$ direction. The waveguide with
the larger longitudinal lattice constant $a_c=440$ nm forms the
nanocavity closed by two surrounding mirror waveguides with
smaller lattice constant $a_m=410$ nm. The target air hole-radius
$r/ a_m$ is 0.293. The semiconductor is etched using Inductive
Coupled Plasma \cite{Talneau2008, Lee2008}, followed by wet
etching and supercritical drying in order to leave the membrane
suspended in air.

\begin{figure}[!h]
\includegraphics[width=8.4 cm]{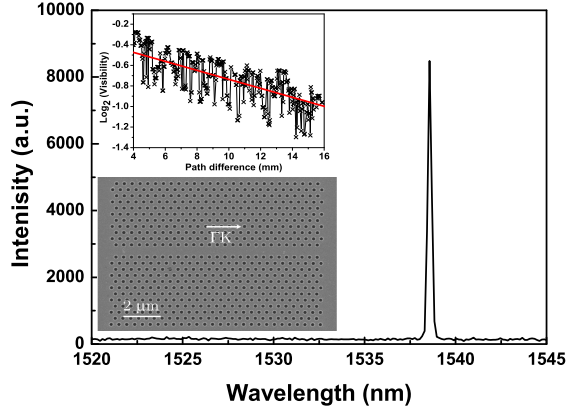}
\caption{\label{fig:Spectrum} Typical microphotoluminescence
spectrum obtained under CW excitation from the double
heterostructure cavity exhibiting a resonant peak at a wavelength
of 1538 nm. Inset (above): Interferogram contrast of the emitted
light decaying exponentially (gray stars and black solid line:
experimental data and red solid line: exponential fit). The
excitation power after the microscope objective is 385 $\mu$W.
Inset (below) : top view of the photonic crystal cavity.}
\end{figure}

The samples are studied at room temperature. Optical excitation is
provided by either a continuous wave (CW) 532 nm Nd:YAG laser or a
pulsed Ti:Sa laser emitting at 840 nm with a repetition rate of 80
MHz and a 5 ps pulse width. The pumping laser is focused to a 5
$\mu$m spot on the sample by a microscope objective (Numerical
Aperture = 0.4). The luminescence is collected by the same
microscope objective and separated from the pumping laser by means
of a dichroic mirror and an antireflection coated silicon filter.
The spontaneous emission is spectrally dispersed by a 0.5 m
spectrometer with a spectral resolution of 0.15 nm and detected by
a cooled InGaAs photodiodes array (Roper Scientific). Emission
linewidths smaller than 0.15 nm are measured by interferometric
analysis of the emitted light on a Michelson interferometer added
in the photoluminescence path. The linewidth is inferred after
fitting the interference fringes envelope by the Fourier transform
of a Lorentzian function (see Fig. \ref{fig:Spectrum}). Lifetime
measurements under pulsed excitation are obtained by means of a
superconducting single photon counter (SSPD-Scontel) with a time
resolution of $\sqrt{2}\sigma=70$ ps, a quantum efficiency of 3\%
at 1.55 $\mu$m and dark count rates lower than 30 counts.s$^{-1}$.
The histograms of the time intervals between the detection of a
photon emitted by the sample after one excitation pulse and the
next laser excitation pulse are recorded on a Lecroy 8620A
oscilloscope. Before being detected by the superconducting
detector, the emission light is filtered by a tunable filter
(Santec, 0.4 nm bandwidth, 20 dB extinction rate).

\begin{figure}[!h]
\includegraphics[width=8.4 cm]{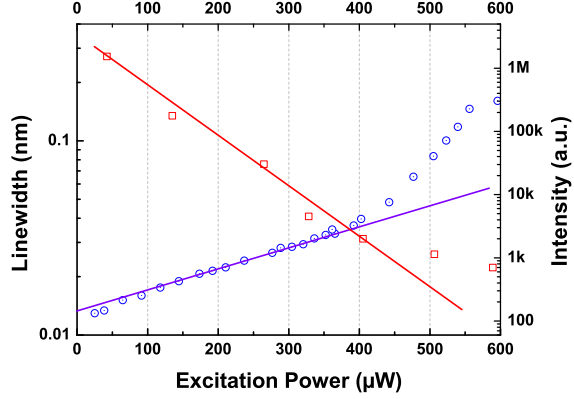}
\caption{\label{fig:Q} (Left axis) Red squares: Linewidth of the
cavity mode as function of the excitation power. (Right axis) Blue
circles: Emission output intensity as function of the excitation
power. Solid lines are guides for the eye. Onset of lasing can be
observed, for excitation powers above 385 $\mu$W.}
\end{figure}

In order to deduce the bare quality factor $Q_{cav}$ of the cavity
mode, we performed linewidth measurements as a function of the
pump power under continuous excitation. Figure \ref{fig:Q}
represents the evolution of the linewidth (left scale) and the
integrated output intensity (right scale) as function of the pump
power. The linewidth exponentially decreases up to a pump power of
385 $\mu$W. Above this excitation power, the linewidth decrease is
slower and the output power increases non-linearly, probably
indicating the onset of lasing. From linewidth measurements at an
excitation power of 385 $\mu$W corresponding to the transparency
of the gain medium, we extract a cavity quality factor $Q_{cav}$
of 44000.

\begin{figure}[!h]
\includegraphics[width=8.4 cm]{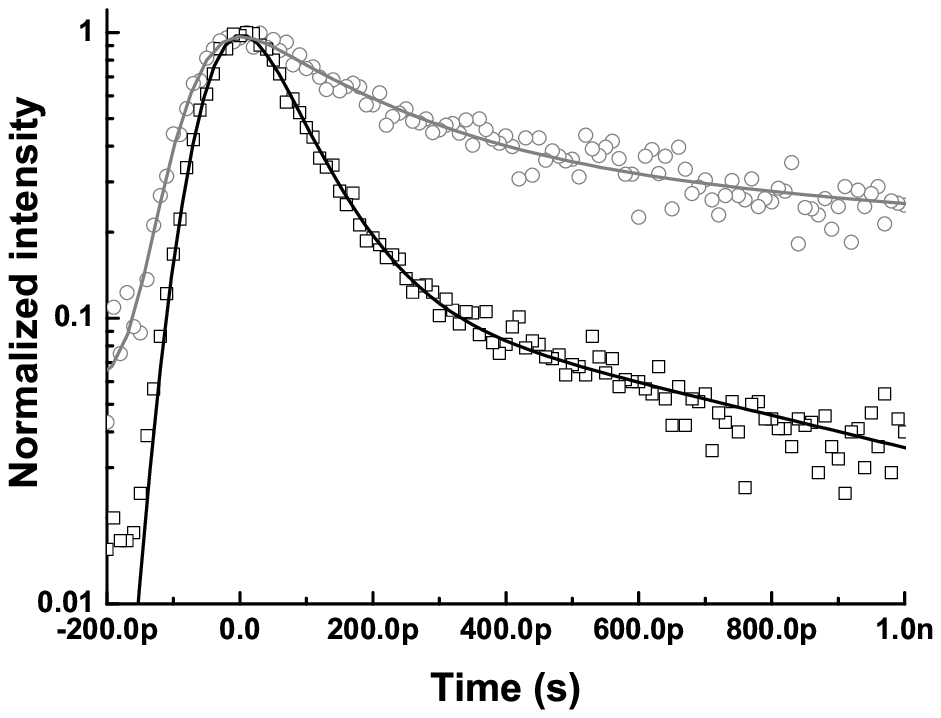}
\caption{\label{fig:Purcell} Lifetime measurements on ensemble of
quantum dots over a spectral window of 0.8 nm centered at
$\lambda=$ 1538 nm. Empty (resp. full) circles represent the
lifetime measured on dots off (resp. on) resonance with the cavity
mode. Dotted line and solid line are fits obtained by use of
equation \ref{fit} for dots respectively off and on resonance.}
\end{figure}
In order to quantify the acceleration of the quantum dots
spontaneous emission rate induced by the cavity, we perform
time-resolved measurements under pulsed excitation. The sample is
excited with 18 $\mu$W average power. Such pulsed excitation power
corresponds to the quantum dots transparency. The measured
lifetimes of the quantum dots respectively off resonance (in the
unprocessed sample) and on resonance with the cavity mode are
shown on Fig. \ref{fig:Purcell}. Both decay curves display a
double exponential decay. We attribute the fast decay to the
emission of excited states (such as an electron-hole pair on the p
shell) of lower energy quantum dots and the long decay time to the
emission of quantum dots ground state \cite{Hostein2008}. The
double exponential feature disappears at low power and low
temperature but is always observed at room temperature for any
excitation power due to thermal population of the p-shell. These
decay times $\tau_{j}^{i}$ ($j=0$ or $cav$ for dots respectively
off and on resonance with the cavity mode; $i=f, l$ for fast and
long decay respectively) can be inferred from fits by
bi-exponential decay curves taking in account the timing
resolution of the detector as follows:
\begin{equation}
I(t)=\sum_{i=s,l} a_i
e^{\frac{\sigma^2}{2(\tau_{j}^i)^2}-\frac{t}{\tau_{j}^i}}\left(1-Erf\left(
\frac{\sigma^2-t\tau_{j}^i}{\sqrt{2}\sigma\tau_{j}^i}\right)\right)
\label{fit}
\end{equation}
\begin{table}
    \centering
        \begin{tabular}{|c|c|c|c|}
        \hline
            & Unprocessed   & Cavity mode &  $\tau_{cav}^{i}/\tau_{0}^{i}$\\
    & sample ($j=0$) &  ($j=cav$) &  \\
            \hline
            Long decay ($i=l$) & 2.14 $\pm$ 0.28 ns & 0.77 $\pm$ 0.06 ns & 2.8 $\pm$ 0.6\\
             \hline
            Fast decay ($i=f$) & 0.20 $\pm$ 0.02 ns & 0.07 $\pm$ 0.03 ns & 2.6 $\pm$ 0.3\\
                  \hline
        \end{tabular}
        \caption{\label{Tab:Purcell} Measured spontaneous emission decay times for the dots in the unprocessed sample and on resonance with the cavity mode.}
\end{table}
The inferred values of the different decay times are summarized in
table \ref{Tab:Purcell}. The ratio $\tau_{cav}^{i}/\tau_{0}^{i}$
between the observed spontaneous emission decay times in the
cavity mode and in unprocessed sample for both long and short
decays allows us to determine the minimal spontaneous emission
enhancement $F$ induced by the cavity, which is of the order of
2.8.  In our experiment, the cavity linewidth is smaller than the
homogeneous linewidth $\delta \lambda_{em}$ of the quantum dot
optical transition at room temperature. The Purcell factor is
hence limited by the quantum dot linewidth and not the cavity
linewidth \cite{vanExter1996, Xu2000}. We can define an equivalent
quantum dot quality factor $Q_{em}$ as the ratio $\lambda/\delta
\lambda_{em}$. In this context, the measured spontaneous emission
enhancement factor on ensemble of quantum dots is equal to:
\begin{eqnarray}
F&=& \frac{1}{2}\frac{\int_{S} |\vec{E}(x, y)|^4 dxdy}{|\vec{E}(x, y)|_{max}^2\int_{S} |\vec{E}(x, y)|^2 dxdy} F_p\\
\mbox{with }F_p &=& \frac{3}{4\pi^2}\frac{Q_{em}
(\lambda/n)^3}{V_{eff}}
\end{eqnarray}
The first term $1/2$ in the expression of $F$ results from the
random orientation of the optical emitting dipoles in the quantum
dot plane $S$ and the second term accounts for the random spatial
distribution of the dots in the central plane of the membrane,
estimated  to 0.17 from our FDTD simulations. The last term $F_p$
is the maximal Purcell factor achieved for a single spatially and
spectrally resonant quantum dot with an optical dipole collinear
to the electric field $\vec{E}$ in the cavity central plane. The
effective volume of the cavity mode has been estimated by FDTD
simulations to V$_{eff}=1.2(\lambda/n)^3$. Consequently,
$F=Q_{em}/186$, which corresponds to a $Q_{em}$ of the order of
500 regarding our measured $F$ values of 2.7. This value of
$Q_{em}$ corresponds to a $\delta \lambda_{em}$ of the order of 3
nm at 300 K, which is rather small compared to InAs/GaAs quantum
dots, but may arise from the large dimensions of the dots
\cite{Besombes2001} (of the order of 8 monolayers height and 40 nm
large \cite{Michon2008}). Moreover, this value of $Q_{em}$ indicates that the spontaneous emission enhancement for such dots
at room temperature would be at most equal to $1/2 \times F_p
\simeq$ 30.

In conclusion, we demonstrated spontaneous emission enhancement
from ensemble of InAsP quantum dots emitting in the
telecommunication C-band at room temperature. This spontaneous
emission exaltation results mainly from the coupling of dots in a
small mode volume double heterostructure cavity, which displays a
single mode with a quality factor of 44000. The enhancement by a
factor of the order of three of the spontaneous emission dynamics
is limited by the homogeneous linewidth of the quantum dots at
room temperature.
The authors acknowledge financial support from Conseil r\'egional
d'Ile-de-France under CRYPHO project.


\end{document}